\documentclass{appolb}
\usepackage{graphicx}
\usepackage{amsmath}
\usepackage{amssymb}
\usepackage{xspace}
\usepackage{cite}
\newcommand{\GeV}{\ensuremath{\,\mathrm{GeV}}\xspace}
\newcommand{\TeV}{\ensuremath{\,\mathrm{TeV}}\xspace}
\newcommand{\nLO}{\nbar \text{LO}\xspace}
\newcommand{\nNLO}{\nbar \text{NLO}\xspace}

\newcommand{\nbar}{\ensuremath{\bar{n}}}
\newcommand{\jet}{\ensuremath{\,\mathrm{jet}}\xspace}
\newcommand{\LS}{{\ensuremath{\,\mathrm{LS}}\xspace}}
\newcommand{\squeezeup}{\vspace*{-1.5mm}}
\newcommand{\squeezeupa}{\vspace*{-3.5mm}}
\newcommand{\etj}{\sum_{\text{jets}}\text{E}_{\text{T},i}}
\newcommand{\etew}{\sum_{W,Z}\text{E}_{\text{T},i}}
\begin{document}
\title{%
\vspace*{-4cm}%
  \begin{minipage}[t!]{1.0\linewidth}
    \begin{flushright}
{\small CERN-PH-TH-2015-241,~FT-UV-15-1103,~IFIC/15-86,~IFJPAN-IV-2015-16,~KA-TH-21-2015}
    \end{flushright}
  \end{minipage}\\\vspace*{1.7cm}\\
Di-boson Production beyond NLO QCD and Anomalous Couplings%
\thanks{Presented by F.Campanario at the International Conference of Theoretical Physics ``Matter to the Deepest'', Ustro\'n (Poland).}%
}
\author{F.~Campanario, M.~Rauch, R.~Roth, D.~Zeppenfeld and
\address{Institute for Theoretical Physics, Karlsruhe Institute of
                Technology (KIT), Germany}
\\~\\
{S.~Sapeta}
\address{CERN PH-TH, CH-1211, Geneva 23, Switzerland}
\vspace*{-0.4cm}
\address{The H.\ Niewodnicza\'nski Institute of Nuclear Physics PAN,
  Radzikowskiego 152, 31-342 Krak\'ow, Poland} 
}
\maketitle
\squeezeupa
\begin{abstract}
In these proceedings, we review results for several di-boson production
processes beyond NLO QCD at high transverse momenta using the VBFNLO
Monte-Carlo program together with the LOOPSIM method. 
Additionally, we show for the $WZ$ production process
how higher order QCD corrections can resemble anomalous coupling effects.
\end{abstract}
\squeezeup
\squeezeup
\section{Introduction}
Di-boson production processes are important channels to test the
Standard Model~(SM) at the LHC. They have been studied intensively in
the past years both from the
theoretical and the experimental side. As a signal, they are
sensitive to triple gauge boson couplings, and therefore, provide a unique
avenue to quantify deviations from the SM predictions. Furthermore, they
are a background to many SM and beyond standard model analyses.
Due to the large size of the next-to-leading order(NLO) corrections and 
the expected percent precision measurement at the LHC, the theoretical
community has pursued in the last years the task to provide
next-to-next-to-leading order~(NNLO) QCD results. This task has been almost
completed in the last years and exact results at NNLO are known for
most of the processes, not only for total cross
sections~\cite{Grazzini:2013bna}, but
also for differential
distributions~\cite{Catani:2011qz}.

At the same time, due to the large collection of results known at NLO
for processes with different jet multiplicities, a field by its own has
emerged with the aim to merge in a consistent way processes with
different jet multiplicities at NLO. 
In this letter, following the LOOPSIM approach~\cite{Rubin:2010xp}, we
will merge $VV$ and $VV$+jet samples and review results 
at approximate NNLO accuracy for several di-boson
production processes. Furthermore, we will show preliminary results of anomalous couplings
effects in $WZ$ production.%
\squeezeup
\section{Ingredients of the Calculational Setup}
Using the LOOPSIM approach, we merge samples at NLO accuracy, provided
by the VBFNLO Monte Carlo program~\cite{Arnold:2008rz}, for several $VV$ and $VV$+jet
production processes. The merged sample is simultaneously accurate at
NLO for both the $VV$ and $VV$+jet sample and provides results at NNLO
accuracy for the $VV$ production process in certain regions of the phase
space since it includes consistently the double-real and virtual-real
contributions to the $VV$ NNLO contributions, simulating in a unitarity
approach, the missing two loop corrections, such that by construction
the merged sample is infrared finite. Thus the sample consistently
includes all the new phase space regions opening up first at NNLO,
including the double soft and collinear emission of the weak bosons,
which leads to numerically large logarithms of the form $\log(p_t^2/M_W^2)$ and
therefore to potentially large NNLO corrections. Furthermore, it
includes consistently the new partonic sub-processes opening up at NNLO,
in this case, $gg$ and $qq$ initiated processes. Thus in regions of
phase space where the LO kinematics are not dominant and therefore the
missing finite piece of the two loop corrections is suppressed, like in
inclusive anomalous coupling searches, the merged sample should provide
most of the NNLO contributions. %
%
%
%
\begin{figure}[t]
  \centering
  \includegraphics[width=0.45\columnwidth]{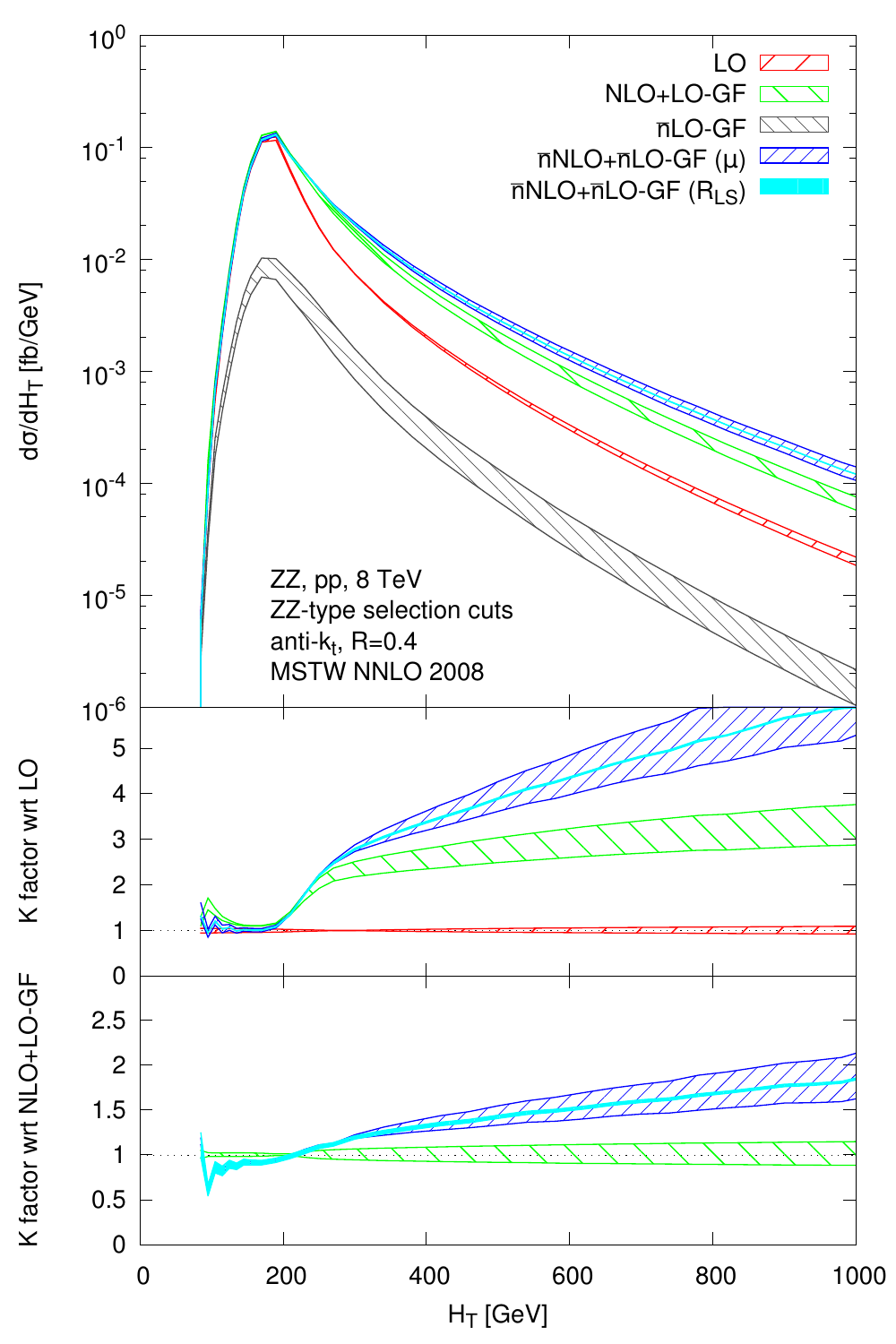}
  \hfill
  \includegraphics[width=0.45\columnwidth]{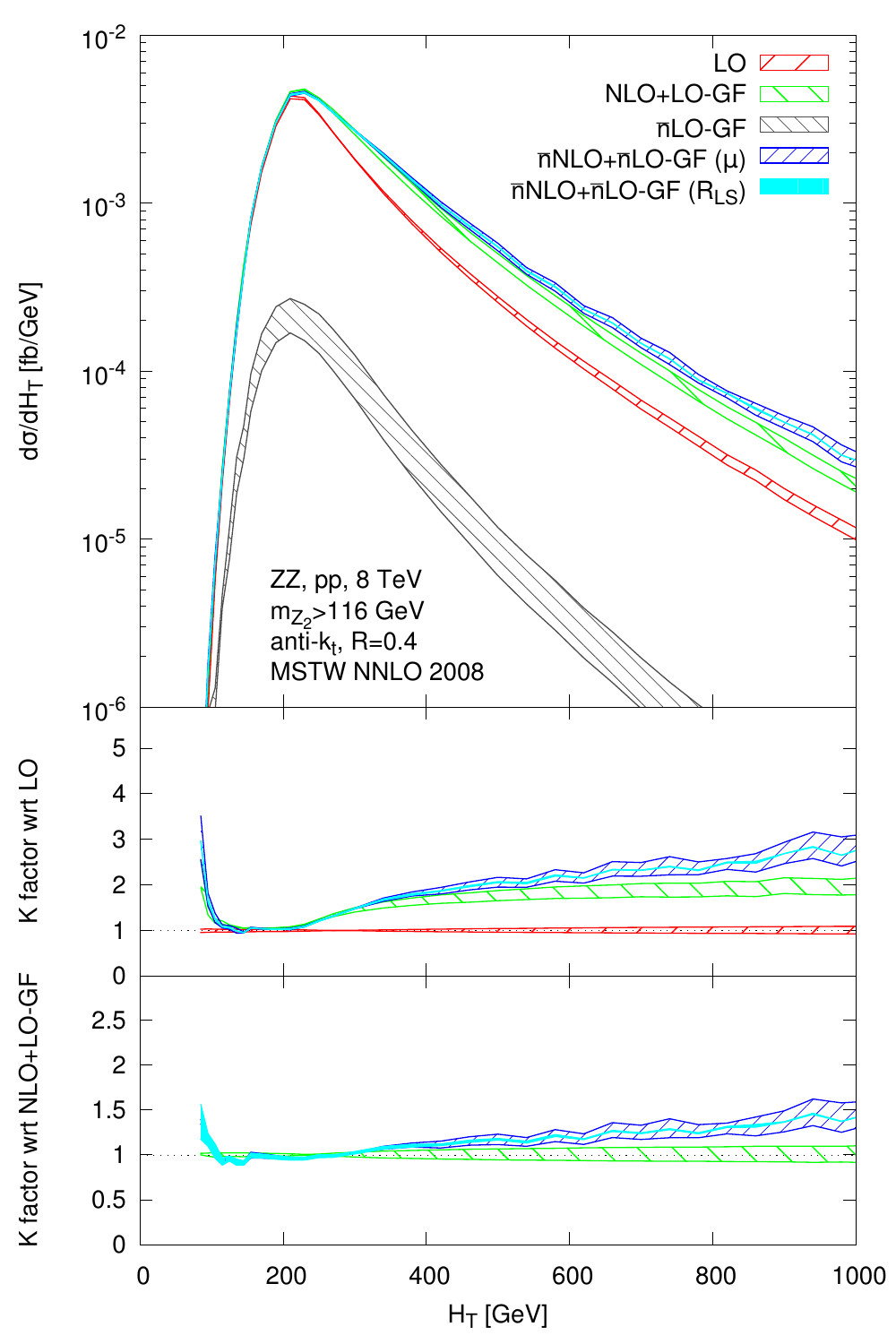}
  \caption{  
  Differential cross sections and K factors for the effective mass observable $H_T$,
  defined in Eq.~(\ref{eq:HT}), for the LHC at $\sqrt{s}=8\, \text{TeV}$.
  The bands correspond to varying $\mu_F=\mu_R$ by factors 1/2 and 2 around the
  central value.
  The solid bands give the
  uncertainty related to the $R_\text{LS}$ parameter varied between 0.5 and 1.5.
  The distribution is a sum of contributions from same-flavor decay
  channels ($4e$ and $4\mu$) and the different-flavor channel ($2e2\mu$)
  in $ZZ$ production.
  }
  \label{fig:HT-standard}
\end{figure}
\begin{figure}[t]
  \centering
  \includegraphics[width=0.49\columnwidth]{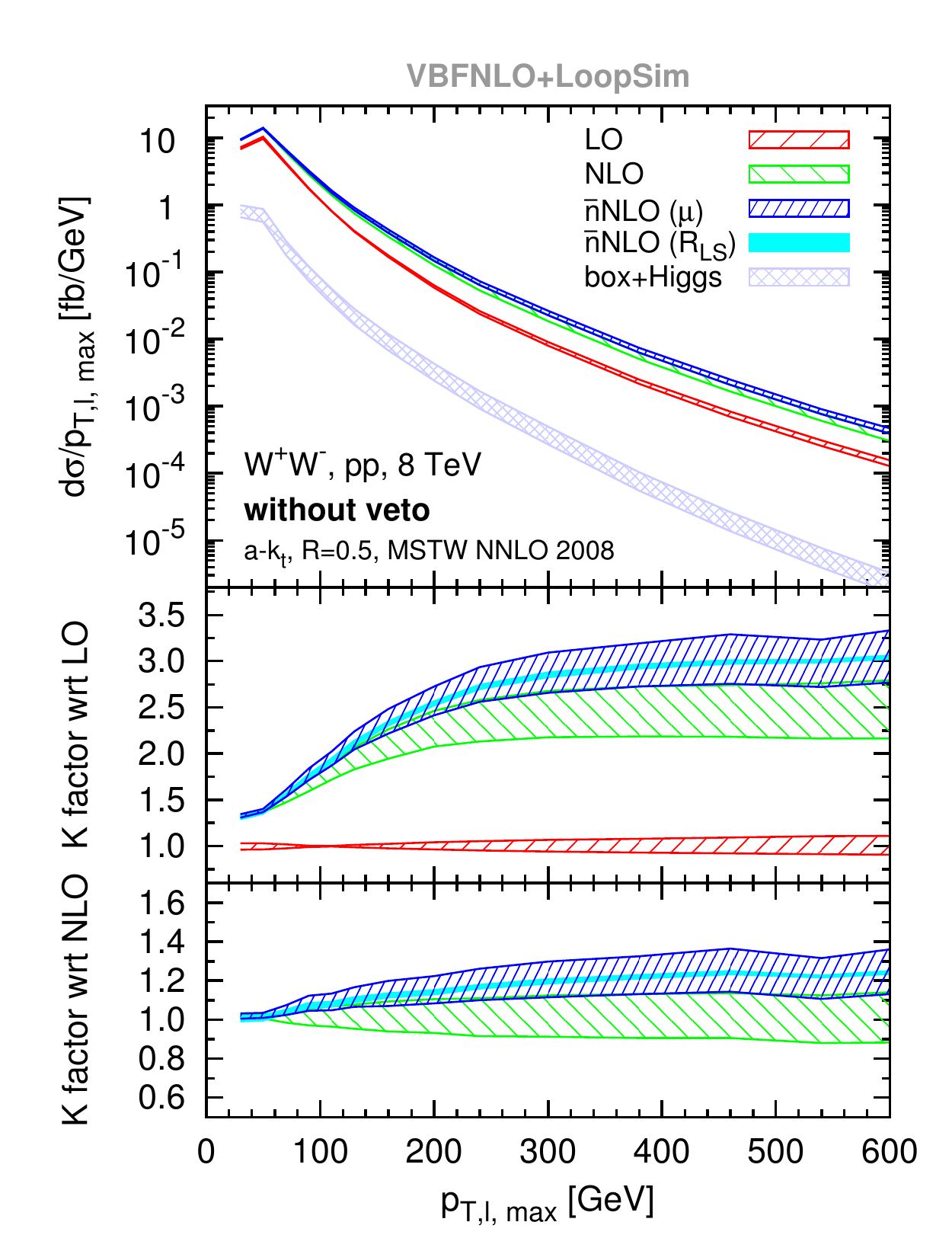}
  \hfill
  \includegraphics[width=0.49\columnwidth]{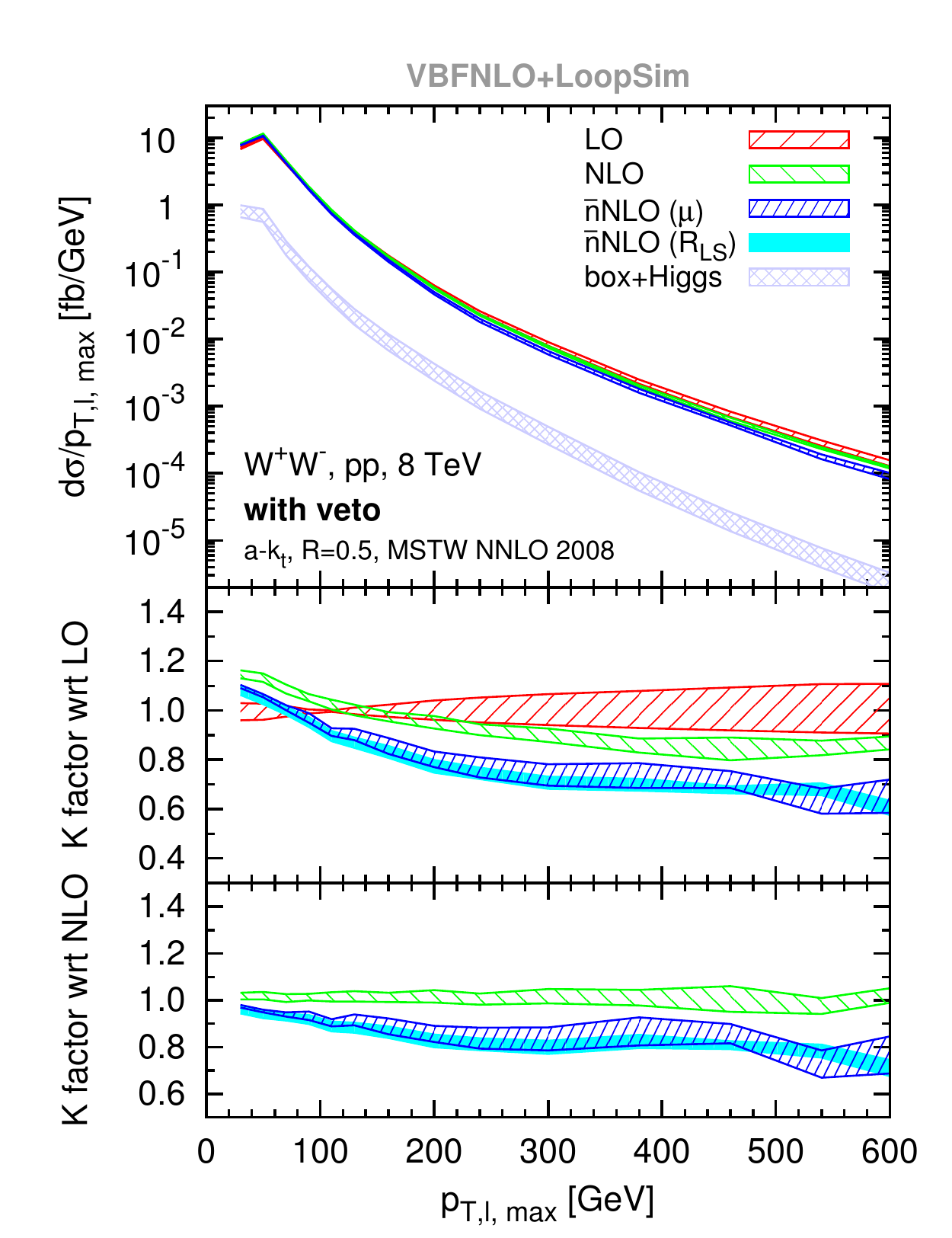}
  \caption{Differential cross sections and K-factors for the $p_T$ of the hardest lepton
  at $\sqrt{s}=8\, \text{TeV}$ without (left) and with
  jet veto (right). Bands are defined as in Fig.~\ref{fig:HT-standard}.
  We include the channels  $e^+  \nu_e   e^-  \bar{\nu}_e$,
  $\mu^+\nu_\mu \mu^-\bar{\nu}_\mu$, 
  $e^+  \nu_e   \mu^-\bar{\nu}_\mu$ and
  $\mu^+\nu_\mu e^-  \bar{\nu}_e$.
  The contribution from the gluon-fusion box and Higgs diagrams is
  included in the NLO and \nNLO curves.
  The left panels correspond to the inclusive sample, while the results shown in
  the right panel were obtained with vetoing events containing jets
  which fulfill the criteria $p_{T,\jet} > 30 \GeV$ and $|\eta_{\jet}| < 4.7$.
   }
  \label{fig:pTWW-standard}
\end{figure}
\squeezeup
\section{SM predictions} 
In the following, results for $ZZ$ and $WW$ production are given at
\nNLO~\footnote{We use $\nbar$ to refer to our approximated results.}. They were studied in Ref.~\cite{Campanario:2015nha} and
Ref.~\cite{Campanario:2013wta}, respectively. The input parameters and a
detailed description of the analysis can be found there. 
We take into account the leptonic decay of the weak bosons including all
off-shell and spin correlation effects. However, we refer to the processes by
the on-shell production mode for simplicity.

%

Independently of the order of a prediction, we used the NNLO
MSTW2008~\cite{Martin:2009iq} PDF set with
$\alpha_s(m_Z)= 0.11707$. As central value for the factorization
and renormalization scales we chose %
$\mu_{F,R}$= $\frac12 \left( \right.$
$\sum  p_{T,\text{partons}} + $
$\sqrt{p_{T,V_1}^2+m_{V_1}^2}+$
$\sqrt{p_{T,V_2}^2+m_{V_2}^2} 
\left. \right) \,,$
%
where $p_{T,V_i}$ and $m_{V_i}$ are the transverse momenta and invariant
masses of the decaying vector bosons, respectively. 
The scale uncertainty is obtained by varying simultaneously the factorization
and renormalization scale by a factor two around the central scale.
Additionally, to assess the uncertainties associated with the recombination
method used by LOOPSIM, we show the uncertainty bands associated with 
variations of the clustering radius, $R_{\text{LS}}$, of $\pm 0.5$
around the central value $R_{\text{LS}}=1$. $R_{\text{LS}}$ is used
in LOOPSIM to establish the sequence of emissions, which is used later on to identify the Born
type particles ($WZ,Vj$ or $jj$) of the event.

Fig.~\ref{fig:HT-standard}, for $ZZ$ production, shows the differential distribution for the effective mass,
$H_T$, defined as a scalar sum of transverse momenta of leptons and
jets \squeezeup
\begin{equation}
H_{T} = \sum  p_{T,\text{jets}} + \sum  p_{T,l}
\,.
\label{eq:HT}
\end{equation} 
The set of cuts closely follows the ATLAS~\cite{Aad:2012awa} analysis 
for inclusive searches and is described in detail in Ref.~\cite{Campanario:2015nha}. 
In the left panel, we require  that the invariant masses of the reconstructed $Z$
bosons satisfy the cut $66 \GeV < m_{\text{inv},Z_i} < 116 \GeV$ and
label  the pair closer (further) to the on-shell value $m_Z$ as
$Z_{1(2)}$. %
One can observe the large
\nNLO contributions, which clearly exceed the scale uncertainties, and
the small LOOPSIM uncertainty provided by the $R_{LS}$ variation. The
origin of the size of the corrections is well understood and is due to
the sensitivity of this observable to additional jet radiation, leading to enhanced
logarithms of the form $\log(p_{T,\text{jet}}^2/M_Z^2)$.  On
the right, one observes smaller corrections once we reduce the size
of the appearing logs by imposing $m_{\text{inv},Z_2}>116 \GeV$. The
plots also show results at \nLO
for the gluon loop-induced contributions, which formally contribute at
NNNLO and use the amplitudes of Ref.~\cite{Campanario:2012bh}.

In Fig.~\ref{fig:pTWW-standard}, we show the differential distribution
for the hardest lepton for $WW$ production with (left) and without
(right) applying a fixed jet veto. To a large extent, our cuts match
the ones by the CMS experiment in Ref.~\cite{Chatrchyan:2013oev}.
The loop-squared gluon-fusion box and Higgs
contributions are separately shown. On the left, one can see that the
\nNLO corrections are of the order of $30\%$ and beyond the NLO scale
uncertainties. On the right, we show results for the vetoed
contributions to point out that exclusive samples are subject to
potentially large negative Sudakov corrections and to show that the
entire $\nNLO$-vetoed result has larger scale uncertainties than the
NLO-vetoed curves. This reveals, partially, accidental cancellation
happening at NLO.  However, as discussed in
Ref.~\cite{Campanario:2013wta}, jet-vetoed exclusive
samples are potentially subject to further corrections from the
constant term of 2-loop diagrams which are not accounted for by the
$R_\LS$ uncertainty band.%
\begin{figure}[t]
  \centering
  \includegraphics[width=0.49\columnwidth]{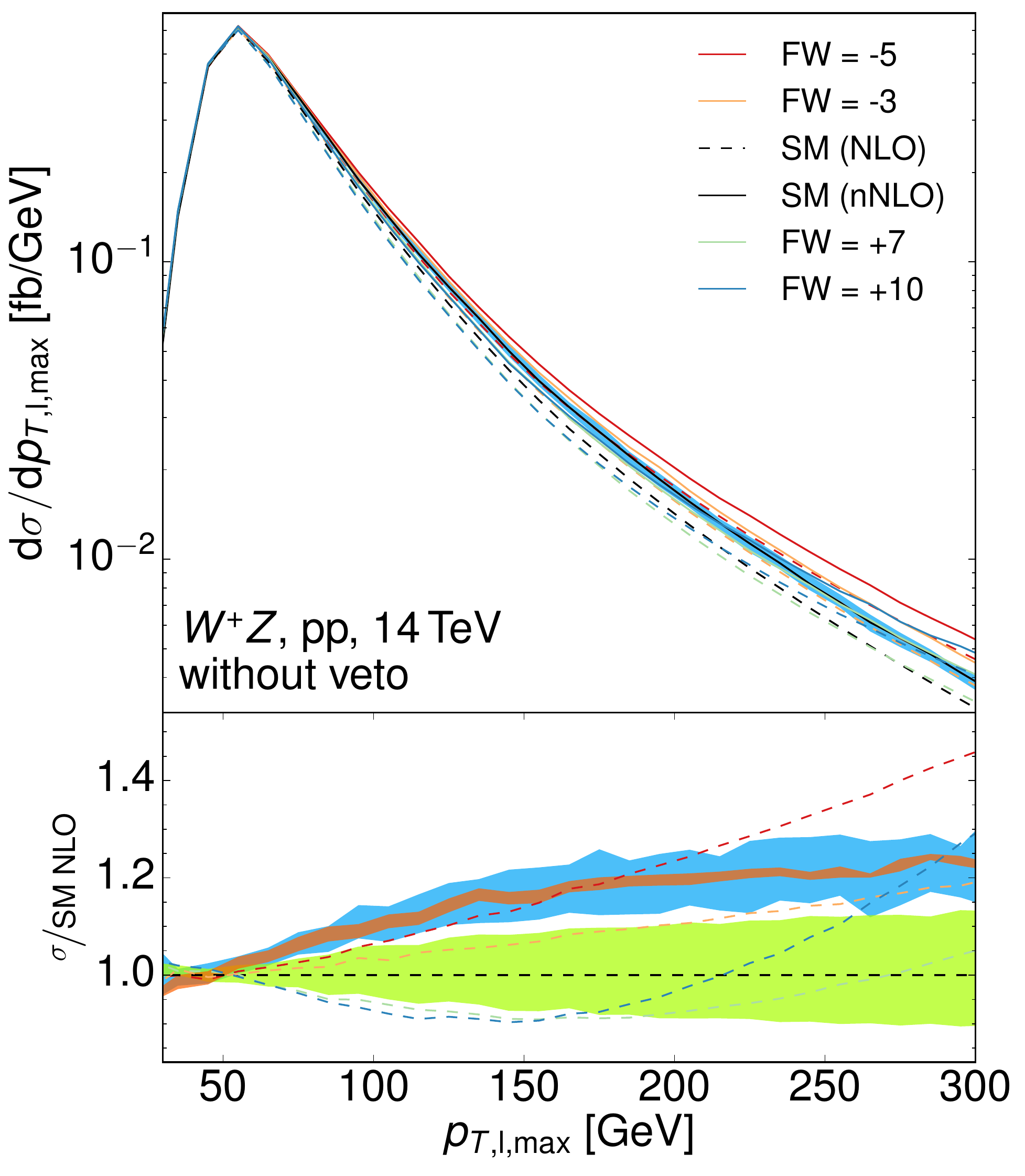}
  \includegraphics[width=0.49\columnwidth]{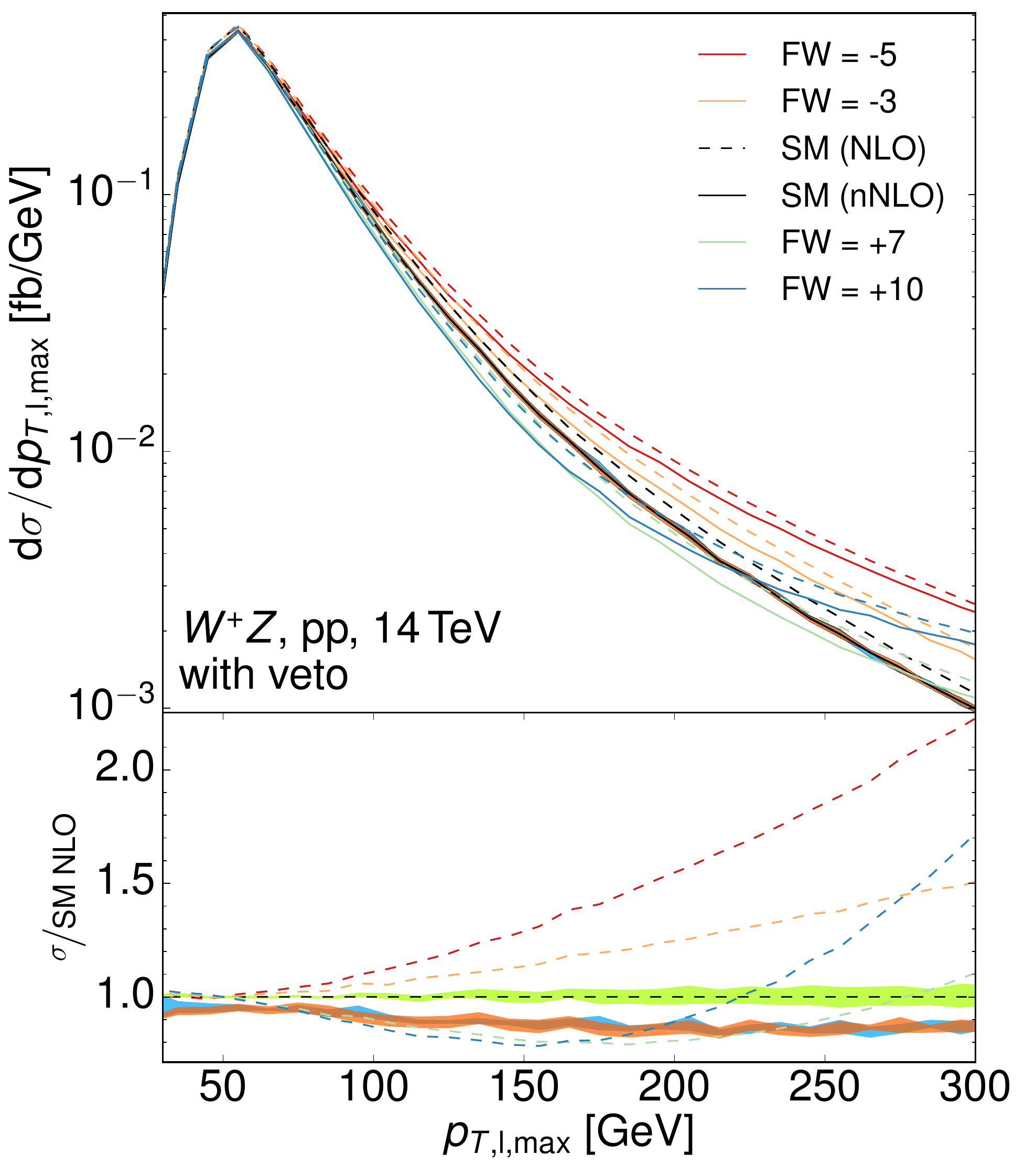}
  \caption{
  Differential cross sections and K-factors for the $p_T$ of the hardest lepton
  for the LHC at $\sqrt{s}=13\, \text{TeV}$ without (left) and with
  a dynamical jet veto (right) for different values of the anomalous
  coupling parameter $F_w$ (in $\TeV^{-2}$). 
  The light-grey (green) and grey (blue) bands correspond, respectively, to the SM NLO
  and SM \nNLO contributions varying $\mu_F=\mu_R$ by factors 1/2 and 2 around the
  central value. The dark-grey (orange) band correspond to the SM \nNLO
  uncertainty related to the $R_\text{LS}$ parameter varied between 0.5 and 1.5.    
Dashed and solid lines refer to NLO and \nNLO, respectively.
   }
  \label{fig:AC}
\end{figure}
\squeezeup
\section{Anomalous Couplings}
In the following, we show how higher order corrections can fake
anomalous couplings~(AC) effects for $WZ$ production. We closely follow the
setup defined in Ref.~\cite{Campanario:2012fk} and use the amplitudes
from Ref.~\cite{Campanario:2010xn}.  On the left plot of
Fig.~\ref{fig:AC}, we present the SM predictions for lepton $p_T$
distributions with a finite anomalous coupling
parameter, $F_w=f_W/\Lambda^2$, corresponding to the dimension 6 operator
$ \left(D_\mu
\Phi \right)^\dag \hat{W}^{\mu\nu} \left(D_\nu \Phi \right)$.  
The coupling values used are within the range allowed by the global fit
to present data in Ref.~\cite{Corbett:2012ja}. %
%
We
use a dipole form factor to preserve tree level unitarity, with a
form factor scale derived from unitarity constraints.
One can clearly see in the left plot that higher order QCD contributions
can fake AC effects, if NLO predictions are taken. 
On the right, to increase the sensitivity to AC, we apply a dynamical
veto, $x_{\text{jet}}<0.2$, as described in
Ref.~\cite{Campanario:2014lza} and given by $x_{\text{jet}} = \etj/(\etj+\etew)$.
\squeezeup
\section{Acknowledgments}
Part of this work was performed on the computational resource bwUniCluster funded by
the Ministry of Science, Research and Arts and the Universities of the State
of Baden-W\"urttemberg, Germany, within the framework program bwHPC.
We acknowledge support from the BMBF (``Verbundprojekt
HEP-Theorie'', 05H12VKG) 
and the DFG (GRK 1694).
FC thanks support to the Spanish Government and ERDF funds from the
European Commission (FPA2014-53631-C2-1-P and FPA2014-54459-P). %
\squeezeup

\end{document}